# ChatGPT's financial discrimination between rich and poor – misaligned with human behavior and expectations.

Dmitri Bershadskyy[1*], Florian E. Sachs[2], Joachim Weimann[1]

[1] Faculty of Economics and Management, Otto-von-Guericke University Magdeburg, Germany

[2] Faculty of Management, Economics and Social Sciences, University of Cologne, Germany

* Corresponding author: dmitri.bershadskyy@ovgu.de, ORCID: 0000-0002-9856-1548

## Abstract

ChatGPT disrupted the application of machine-learning methods and drastically reduced the usage barrier. Chatbots are now widely used in a lot of different situations. They provide advice, assist in writing source code, or assess and summarize information from various sources. However, their scope is not only limited to aiding humans; they can also be used to take on tasks like negotiating or bargaining. To understand the implications of Chatbot usage on bargaining situations, we conduct a laboratory experiment with the ultimatum game. In the ultimatum game, two human players interact: The receiver decides on accepting or rejecting a monetary offer from the proposer. To shed light on the new bargaining situation, we let ChatGPT provide an offer to a human player. In the novel design, we vary the wealth of the receivers. Our results indicate that humans have the same beliefs about other humans and chatbots. However, our results contradict these beliefs in an important point: Humans favor poor receivers as correctly anticipated by the humans, but ChatGPT favors rich receivers which the humans did not expect to happen. These results imply that ChatGPT's answers are not aligned with those of humans and that humans do not anticipate this difference.

# Introduction

ChatGPT became an initial point for the popular application of large language models (LLMs). In contrast to prior models, GPT-3 facilitates even naïve and inexperienced users to receive (sometimes only apparently) sophisticated answers to almost any type of question. While LLMs as such are very powerful in a variety of contexts (Dwivedi et al., 2023), a naïve utilization of this tool may cause some problems. Foremost, research hints at a variety of ethical concerns (Stahl & Eke, 2024). Further, the tools sometimes provide false information while making it look like a highly sophisticated answer. At least, there is already anecdotal evidence pointing out that the use of LLMs benefits users with subpar performances in easy tasks but hampers them to build up knowledge for complex tasks. While LLMs often ask the users to check the quality of the provided answer, it is unclear and seems to be doubted whether users follow such advice. If humans take over advice from LLMs in questions where other humans are affected, it is worth investigating whether such suggestions are aligned with their preferences and how other humans react to an LLM making the call.

As of now, there are several applications of ChatGPT in the economics context. It remains unclear which of these will come into practice in the real world. However, we consider resource allocation problems to be something where LLM can be applied very simply (e.g., gift making (Kirshner, 2024)). ChatGPT provides this type of managerial advice to monetary allocation problems without any notes of caution. However, for making allocation decisions, fairness preferences are essential (Bolton & Ockenfels, 2000; Charness & Rabin, 2002; Fehr & Schmidt, 1999). Until now, it is unclear whether (and what) fairness preferences can be gathered within an LLM. While some argue that LLMs can collect sufficient information to approximate such preferences, this is far from certain and requires tests.

Further, even in the case where LLMs collect some type of fairness preferences, whenever it comes to bargaining about resources, there is a second element that becomes important – expectations about the other side. Therefore, in certain bargaining scenarios, it is beneficial to know the fairness preferences of the other side even if it is not human.

In our experiment, we aim to address all three of these issues. In precise, first, we assess how humans react in a simple bargaining situation, i.e., the Ultimatum Game (UG), when the proposer is a Chatbot (here specifically ChatGPT). Second, we investigate whether offers from ChatGPT are similar to human offers in a standard laboratory experiment. Note that due to the UG experiment being very famous, ChatGPT should be able to rebuild the most probable results (at least for a student sample which is commonly used in UG experiments). This would not imply anything about depicting human preferences. As human preferences are not easy to depict, we applied a small change in the classical UG involving a certain case of discrimination due to financial endowment. We investigate the consistency of ChatGPT's answers and compare these answers to those of humans facing the same allocation task. Third, we investigate human expectations about the interaction with ChatGPT in this context.

Further, note that from the literature we know that receivers in UG can react differently to the same size of an offer for different reasons. Importantly for our research question, some receivers were ready to accept lower offers when those came from random events instead of intentional human action (Blount, 1995; Bolton et al., 2005). Looking at the acceptance behavior of receivers leads to the question of whether receivers behave towards ChatGPT more like a human or a random event.

In so doing, this article contributes to the emerging literature that uses experimental methods from economics (Kirshner, 2024; Phelps & Russell, 2023), social sciences (Xu et al., 2024), and psychology (Pellert et al., 2024) to better comprehend the behavior of LLMs. Yet, we differ from these by adding two core issues. First, we apply a game that cannot be part of ChatGPT text corpus making it difficult to simply mimic the most observed results. Second, we combine our analysis with the investigation of human expectations about such behavior to better assess potential alignment problems. Further, we differ from other studies that use UG or Dictator Game together with LLMs by using ChatGPT as the proposer and not the receiver as in other studies (Aher et al., 2023; von Schenk et al., 2023).

Our results indicate that humans expect ChatGPT to behave no differently than human proposers in the bargaining situation of an UG. This may be the reason why we surprisingly do not observe differences in the acceptance behavior between human- or ChatGPT-made offers. Further, human proposers tend to give more money to lowly endowed receivers. This contrasts ChatGPT's behavior as it usually proposes more money to highly endowed receivers. Therefore, we shed first light on the problem that arises from wrongly built expectations about the behavior of chatbots.

## Methods

In this section, we, first, describe our variation of the UG and then describe our two-step procedure for its execution in the lab. We did not implement the standard UG (Güth et al., 1982) for the following reasons. The UG is widely known as indicated by more than 6600 citations of the original publication and over 32,000 results on Google Scholar on "Ultimatum Game". Further, it is so simple that there are thousands of other sources (books, blogs, newspapers, etc.) covering the most important results. Therefore, ChatGPT, by giving the most probable answers, should be able to provide similar results as in the lab. Instead, we aimed to use a game, that to the best of our knowledge, has not been played yet. This means ChatGPT could not have used it in its training set.

The classical UG consists of a simple task. The proposer has a certain amount of money that has to be split between him and the receiver. If the receiver accepts the offer, the distribution is paid out. Otherwise, both participants leave empty-handed (Güth et al., 1982). Decades of research on UG indicate canonical findings, e.g., proposers' offers lie mostly between 40% and 50%, offers of such size are almost always accepted, and offers lower than 20% are usually rejected. In total, this implies that proposers do not only consider their payoff but also the one of the receivers. Further aspects influencing the offers are the awareness that the receivers could reject an offer and the proposer's own fairness concerns (Güth & Kocher, 2014).

Similar to the standard UG, in our version, proposers are asked to split 16€ (in integer values) between themselves and the receivers. Yet, we introduce two types of possible receivers: rich and poor. In the context of our experiment, this means that some rich receivers have an initial endowment of 8€ and poor receivers of 4€. The proposers do not have any endowment. The endowment is independent of the UG. In line with the standard UG, the proposers make an offer. If it is accepted, it results in the respective allocation of the 16€ on top of the endowment. If it is rejected, the 16€ are gone – proposers leave empty-handed, and receivers keep their initial endowment. Further, we apply the strategy approach, i.e. we ask the proposer to provide offers for receivers with 4€ and 8€ respectively at the same time, before they learn the true endowment of their matched receiver. This gives us more observations and better control to understand whether and how proposers (human or ChatGPT) discriminate between the two groups. Focusing on the acceptance behavior, we ask the receivers to indicate their minimum acceptable offer (MAO). This approach is in line with prior literature when it is important to investigate the acceptance behavior of receivers between different groups of proposers, as otherwise, the responses of the receivers would always depend on the observed offer of the proposers. This is especially crucial in our case, as the variance in ChatGPT's proposals and, thus, the observable acceptance behavior of the receivers is limited. Now, we explain our implementation of this game for ChatGPT and humans. For our ChatGPT treatments, first, we prompted the instructions for our modified UG experiment. Note, that the instructions (see Supplementary Information) were identical for humans and ChatGPT. We repeated the same prompt 200 times over the classical ChatGPT interface. We did not use API-based queries to ensure that the answers would appear in the same way users would interact with a Chatbot. Then, we calculated the average offers for both types of receivers and implemented them as the offer from the Chatbot in our laboratory experiment. If the offer from ChatGPT is accepted, the money goes to OpenAI. In the experiment, we did not mention ChatGPT and OpenAI but spoke of the company that developed the chatbot. We are aware of research indicating differences in human behavior depending on how the bot is framed and where the money goes (von Schenk et al., 2023). We purposefully intended to have a neutral formulation.

Similar to our ChatGPT treatment, we had human treatments. Participants of the experiment were randomly chosen to be in the ChatGPT or human treatment. In the latter case, also the assignment as proposer or receiver was random. The instructions were read aloud, making the information about the different endowments public knowledge. The instructions also included information on how receivers indicate their acceptance of the proposal. After this stage, the receivers obtain information on whether the proposer they will interact with is a human or a chatbot. Then the game started.

In total, this yields a 2x2 between-subjects design with two types of proposers (Human or Chatbot) and two types of receivers (High and Low Endowment). In all four treatments, we continued the experiment with the following questionnaire. First, participants answered four questions on their expectations about offers, i.e. what do they think Humans/ChatGPT offer on average to receivers with Low/High endowment? Subsequently, they answered four questions on what they think the receivers (High or Low) on average considered to be MAO for Humans and ChatGPT. All these eight questions were financially incentivized for 80% of all participants. For each of the two sets of four questions, subjects were informed that the closer they are to the true values, the higher their chances of winning an additional 1€. We used the first 20% of participants to gather the information for true values, meaning these observations were not incentivized. Comparing between incentivized and non-incentivized answers indicates no significant differences for any of these questions. In the end, we collect demographic data, implement the German Version of the short form of the Big Five Inventory-2 (Rammstedt & John, 2007), and ask items on negative reciprocity and altruism from the global preferences survey (Falk et al., 2018).

The experiment was organized using hroot (Bock et al., 2014) and conducted in oTree (Chen et al., 2016) at a laboratory for experimental research MaXLab at the University of Magdeburg). The average duration was approximately 15 minutes and the average payoff was approximately 10.00 €. The experimental design is approved by the German Association for Experimental Economic Research. The investigation was conducted in accordance with the principles expressed in the Declaration of Helsinki. All subjects gave written informed consent. The study was preregistered at OSF.io (https://osf.io/85yb3/?view_only=23bf8459f7df461d8e890d41f9ee1751).

## Results

In the result section, we will first briefly present descriptive statistics. Then, we divide our results into sections following our preregistered analysis plan. Therefore, second, we investigate whether participants in the experiment are willing to accept offers of a different size when they come from a chatbot as compared to a human. Third, we investigate differences in offers for receivers with high and low endowments. Fourth, we analyze the expectations of humans on what other humans or the chatbot would do. Fifth, we present some exploratory analysis on other collected variables.

For technical reasons, we achieved slightly more observations than preregistered (47): 51-55 observations in the respective treatment. In line with our preregistration, we did not have to exclude any participant. The participants were on average 23.9 years old, of whom 53.7% were male, 46.0% female, and 0.3% non-binary. There are no significant differences in gender or age between the treatments. Concerning the responses from ChatGPT, we obtained the results that ChatGPT discriminates significantly (Wilcoxon Signed Rank test, p=0.0003) between participants with high and low endowments. Figure 1 depicts the details of ChatGPT's offer distributions. Despite identical prompts, there is a wide range of offers from ChatGPT. On average, ChatGPT offered 6.47€ for subjects with low endowment and 6.94€ for those with high endowment. In line with our preregistration, we therefore implemented 6€ as the offer of the chatbot for all receivers in the low treatment and 7€ for those in high treatment. Note that the differences in terms of the medians are even higher.

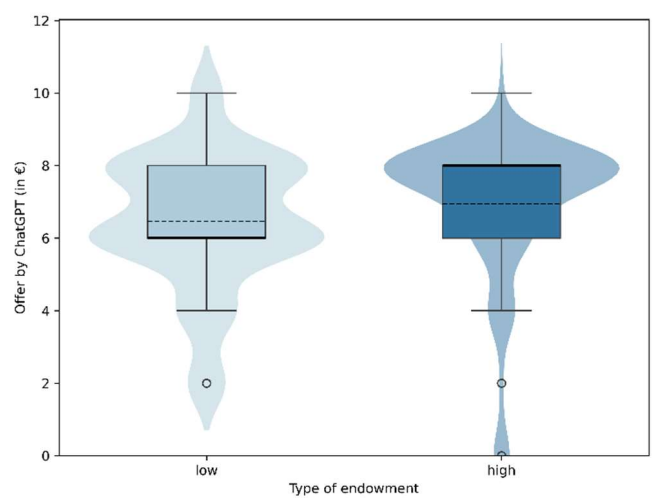

Figure 1. Boxplots for offers from ChatGPT for the different endowment types including median (bold lines) and mean (dotted lines).

Second, we focus on the receivers' MAOs. Concerning the source from where the offer comes (human or chatbot) we do not see any significant differences, neither for the case of low (5.63€ vs. 5.51€) nor the case of high (5.58€ vs. 5.27€) endowment. Next, we investigate whether the MAO depended on the size of the endowment. When comparing the minimum accepted offers from humans we see no significant differences (5.63€ for Low and 5.58€ for High Endowment). The results are similar for humans accepting offers from the chatbot (5.51€ for Low and 5.27 for High Endowment). This means that the participants did not condition their acceptance neither on the source of the offer nor their endowment.

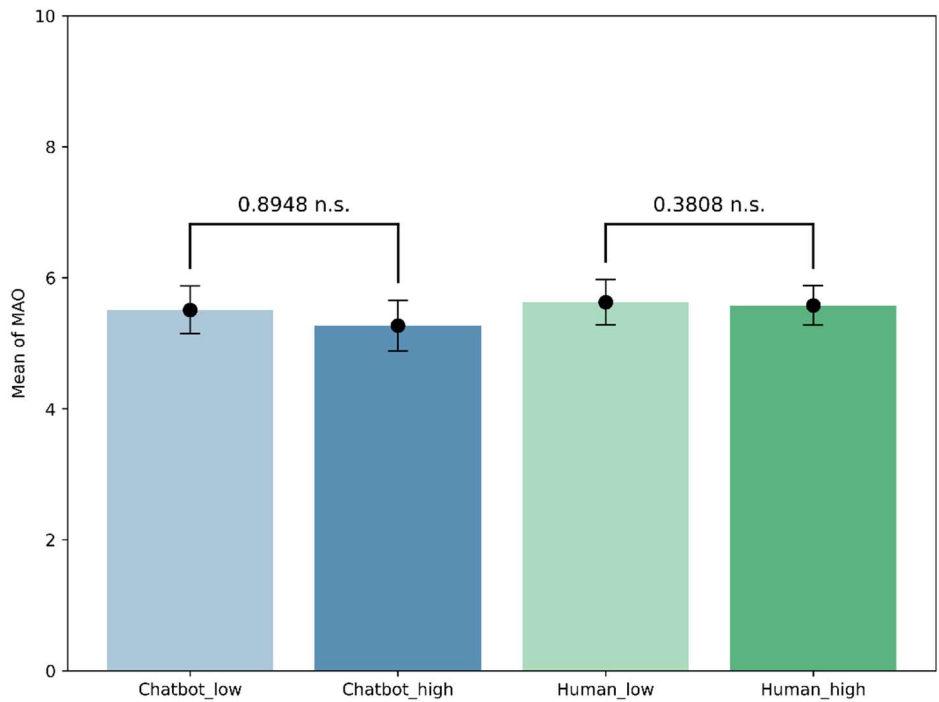

Figure 2. Minimum acceptable offers (MAOs) of all treatments.

Third, we investigate the offers of the proposers. In line with the preregistration, we use the ratio of proposed payoffs for the proposer and receiver to make the proposals more comparable between low and high endowments. Here, a ratio of 1 means that the proposed offer would yield a fair split between the proposer and receiver, taking the receivers' endowment into account. We present the results for all proposals in Figure 3. We see that for the lowly endowed receivers, proposers' offers were on average close to the fair split (ratio=0.95). Yet, for highly endowed receivers, proposers' offers were on average to their disadvantage (ratio=0.77). The difference is highly significant (Wilcoxon Signed Rank test, $p<0.0001$). We, now, investigate the values generated by the chatbot. While the ratio for lowly endowed receivers is similar to humans (0.99), the ratio for highly endowed receivers is 0.65. Again, these differences are highly significant (Wilcoxon Signed Rank test, $p<0.0001$). When comparing the ratios between humans and chatbot for both types of receivers we see no significant differences for the lowly endowed receivers, yet the difference is significant for highly endowed receivers (Wilcoxon Rank Sum test, $p<0.0001$).

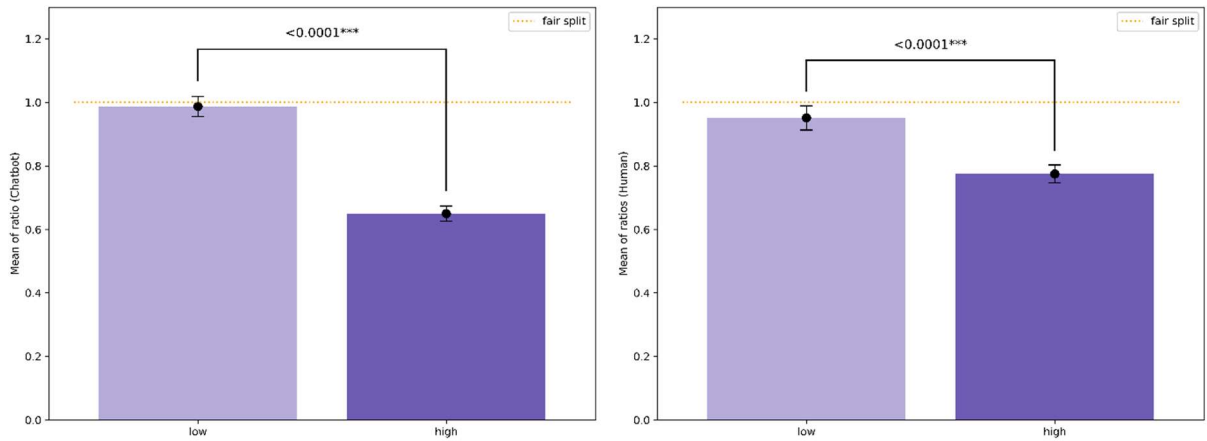

Figure 3. Ratio of proposer and receiver payoffs for chatbot (left) and human (right) proposals. Dotted lines represent the fair split.

Further, we display that these results are robust to taking other metrics. If you are to simply investigate whether the proposers (human or chatbot) discriminate between the two differently endowed types of receivers we observe that humans give on average more (6.60€) to the lowly endowed receivers than highly endowed receivers (5.91€). This contrasts starkly the direction in which ChatGPT discriminates the two groups (6.47€ for low and 6.94€ for high). Figure 4 shows these results.

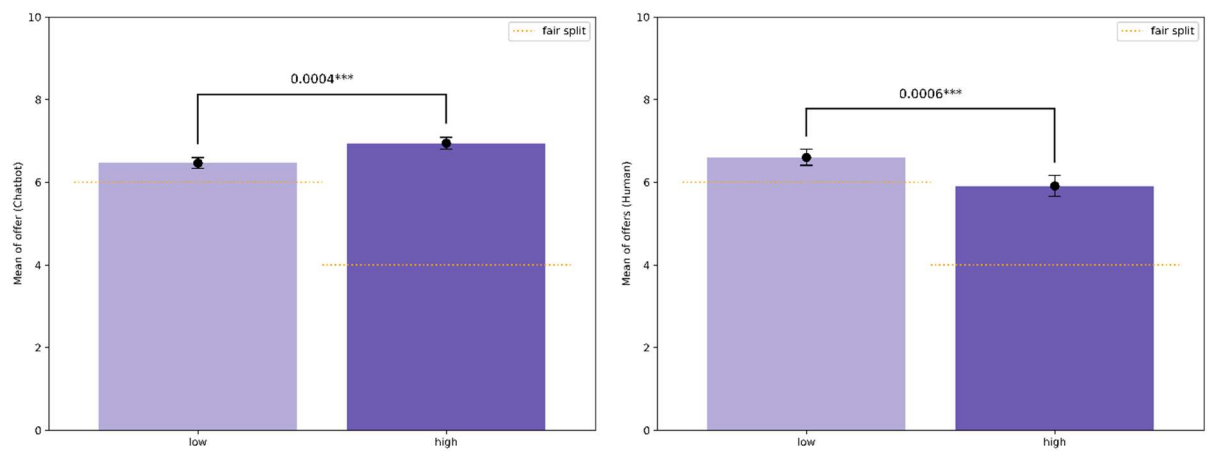

Figure 4. Offers for lowly and highly endowed receivers for chatbot (left) and human (right) proposals. Dotted lines represent the fair split.

Fourth, as in typical bargaining situations, it is important to form correct expectations about the behavior of the other side, we now illustrate whether human participants were able to assess the behavior of others. Doing so we therefore assess whether humans were good at forming expectations about the chatbot. In total, we assessed 8 different types of beliefs. As depicted in Figure 5 (left), the results show that our participants expected that the other human (chatbot) offered on average 6.37€ (6.30€) if the receiver was lowly endowed and 5.48€ (5.33€) if the receiver was highly endowed. The differences are not significant when comparing the values between humans and chatbot. This indicates that on average participants expected the chatbot to behave similarly to other humans. Yet, the differences are highly significant when comparing the respective differences between lowly and highly endowed receivers (both Wilcoxon Signed Rank test, p<0.0001). These results indicate that the participants were able to assess that both humans and chatbot would provide different offers. Yet, the chatbot provided higher offers for receivers who are highly endowed which contrasts human proposers.

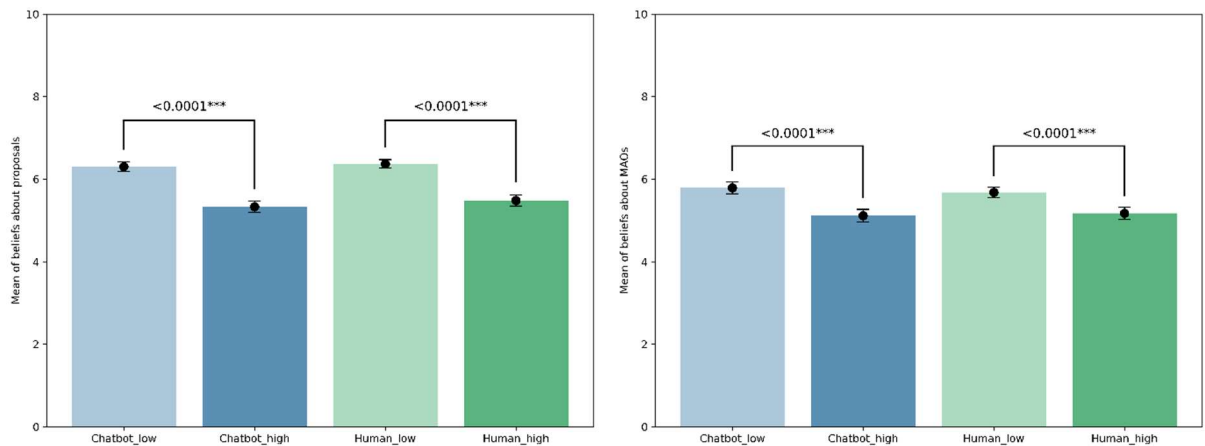

Figure 5. Beliefs about proposals (left) and MAOs (right) for chatbot and human proposers for both receiver endowment types.

Now, we investigate what the participants expected other humans would be ready to accept from humans or chatbot. Figure 5 (right) shows the results. The results indicate that the participants expected receivers with a low endowment to have MAOs of 5.68€ (5.79€) when the other player was human (chatbot). For highly endowed receivers the expectations were for 5.17€ (5.11€) human or chatbot respectively. Again, we do not observe significant differences between MAOs when the offers come from humans vs. chatbot. Yet again we observe strongly significant differences in expectations concerning the endowment of the receivers (both Wilcoxon Signed Rank tests, p<0.0001).

Finally, we provide some regressions and exploratory analysis on other collected variables. In line with our preregistration, we investigated several additional variables for their correlation with our main variables (offers and MAO). We display the correlation matrices in the supplementary materials. In brief, the results indicate that the offers correlate significantly only with the beliefs, meaning that there are no correlations with the Big Five items, questions on negative reciprocity and altruism, or demographics. The results for MAOs are similar, meaning that again there are only meaningful correlations between MAO and the beliefs. These results are depicted in Supplementary Tables S1-S3.

## Discussion

In this section, we first briefly discuss the theoretical predictions of the neoclassical model and then investigate how well the results can be explained by these.

The subgame perfect Nash Equilibrium of the ultimatum game indicates that the proposer should offer the smallest divisible amount of money to the receiver and receivers should accept every offer other than zero. In our experiment, we do not observe any such behavior. The offers of the chatbot and the humans are significantly different from zero. This is a well-known finding for humans which is usually explained by two arguments. First, humans have fairness preferences (Bolton & Ockenfels, 2000; Fehr & Schmidt, 1999), implying a positive utility from a fairer allocation. Second, there may be concerns that the other party may behave "irrationally" and reject a non-zero offer. Therefore, proposers, for strategic reasons, would offer a higher share of the total endowment. In total, while this is known for humans, it leads to the question of whether the same is expected from a chatbot. Since LLMs operate based on a large text body including a lot of information on human behavior in general and behavior in an UG in specific, it is not surprising that ChatGPT's offers are more like those of a typical human and less like those of a homo oeconomicus (as a neo-classical model).

However, what is interesting is the ratio of proposer and receiver payoffs. In the standard UG, the proposers keep between 50%-60%. Translating these values to our measure implies a ratio of 1-1.5 between proposers' and receivers' payoffs. In our variation of the UG, the Nash equilibrium remains the same, yet the observed average ratios were lower than 1. This means that the change of our design introducing endowments for the receiver additionally leads to formerly unobserved behavior. Despite certain similarities in design, this change cannot be fully explained by results from UG with outside options for receivers. In such an experiment, the receivers obtain an alternative amount of money in case they reject the offer (Kahn & Munighan, 1993). Yet, note that when the receiver has an outside option this changes the Nash equilibrium. Importantly, Knez and Camerer (1995)) show that introducing outside options leads to lower acceptance rates as players have more different interpretations of what is a fair offer. This could have occurred in our experiment, too. Most related to our research are UG, with different conversion rates of the allocated chips to money. Respondents with higher conversion rates as in (Kagel et al., 1996) or (Schmitt, 2004) receive higher offers than those with lower rates. Yet, this line of experiments is subject to efficiency concerns since the total group payoff increases the more chips are offered to the player with higher conversion rates. Still, this research supports the idea that proposers try to make offers that will not be rejected by responders instead of solely fairness concerns. This is in line with our observation that proposers discriminated between receivers with high and low endowments.

Concerning ChatGPT's offers, again consider that offers resulting in a fair split were 6€ and 4€. Yet, the observed offers from ChatGPT were different, especially highlighting the differences in the ratio parameter for low and high treatment (0.99 vs. 0.65). This poses the question whether ChatGPT's offers were driven by any type of fairness preferences. If fairness preferences are the dominant aspect for the offers from ChatGPT these values should have been more similar. If, alternatively, ChatGPT actively discriminates between rich and poor, this should be investigated more thoroughly and displayed by the company more transparently. Still, in our case ChatGPT favors rich players.

Yet, what may be even more surprising is the behavior of the receiving humans. Considering typical reasons to reject non-zero offers, we investigate the following: (i) the proposal is considered unfair as initially both persons in the experiment are equal and (ii) the receiver wants to punish/teach the proposer. Now, the question comes up as to whether these reasons are valid in the case of ChatGPT. We know from prior research that receivers do not only assess the monetary value of the offer but also the intention with which it is made (Falk et al., 2003). Looking at experiments where the offer depended on random events instead of intended human action, illustrates that receivers reject these less often displaying lower (by factor 2) MAOs (Blount, 1995; Bolton et al., 2005). We do not observe this behavior. Instead, MAOs for the chatbot in our experiment are indistinguishable from those for humans. This means that receivers consider the chatbot more like a human. Possible explanations may be referred to as some type of anthropomorphism or that humans have some type of hardwired desire to punish intentionally unfair proposals and assume chatbots can have intentions or that the negative feedback from rejection is fed back to the chatbot (even though we did not imply this at any time).

Finally, concerning the beliefs it is an important observation that the expectations were very accurate for all six variables that concerned human behavior (four MAOs and two proposals). All average MAOs were in the range between 5.27€ and 5.63€ while the expectations were between 5.11€ and 5.79€. The same is true for actual human proposals: 6.61€ for low and 5.91€ for high as well as the respective expectations of 6.37€ and 5.48€. However, the participants had wrong beliefs about the chatbot. The chatbot discriminated benefitting the higher endowed group (6.94€ vs 6.47€ for the low). Yet, the participants believed in the same proposing behavior as humans and expected the chatbot to favor the poor group. This leads to either questioning the participants' beliefs or ChatGPT's offers. Independent of this, it clearly indicates that ChatGPT's behavior is not aligned with human expectations. Further, ChatGPT's offers were mostly inconsistent and for lowly endowed receivers the offers appear to be even bimodal. Such inconsistent response behavior is further in line with prior research on consistency of e.g., moral advice (Krügel et al., 2023) and can pose additional threats for establishing AI alignment.

In the end, we consider some limitations. First, this refers to the ever-updating process of ChatGPT. It may occur that the observed offers would differ for other versions. We explicitly focused on the currently freely available version as it is the most used one. Second, based on individual experiments, you cannot make generalizable statements on the alignment problem of ChatGPT. Instead, we consider this to be a contribution to a better understanding of the values that ChatGPT displays in its responses to humans. However, as we implemented the strategy approach using MAOs, different offers by ChatGPT would not affect our results.

Future research should focus on the reasons why humans do not differentiate between human- or chatbot-made offers, why chatbots discriminate between different types of humans and whether chatbots can approximate fairness preferences. The literature on the UG can support all these research questions.

## Acknowledgements

This work was supported by the OVGU/FEM Research Data Fund.